# Probes for ultra-sensitive THz nanoscopy

*Curdin Maissen[1,*], Shu Chen[1,*], Elizaveta Nikulina[1], Alexander Govyadinov[2], and Rainer Hillenbrand[3,4]*

[1]CIC nanoGUNE, 20018 Donostia-San Sebastián, Spain
[2]Neaspec GmbH, 85540 Haar, Germany
[3]CIC nanoGUNE and UPV/EHU, 20018 Donostia-San Sebastián, Spain
[4]IKERBASQUE, Basque Foundation for Science, 48013 Bilbao, Spain

[*]*equally contributing authors*
*Corresponding author: r.hillenbrand@nanogune.eu*

**Scattering-type scanning near-field microscopy (s-SNOM) at terahertz (THz) frequencies could become a highly valuable tool for studying a variety of phenomena of both fundamental and applied interest, including mobile carrier excitations or phase transitions in 2D materials or exotic conductors. Applications, however, are strongly challenged by the limited signal-to-noise ratio. One major reason is that standard atomic force microscope (AFM) tips – which have made s-SNOM a highly practical and rapidly emerging tool - provide weak scattering efficiencies at THz frequencies. Here we report a combined experimental and theoretical study of commercial and custom-made AFM tips of different apex diameter and length, in order to understand signal formation in THz s-SNOM and to provide insights for tip optimization. Contrary to common beliefs, we find that AFM tips with large (micrometer-scale) apex diameter can enhance s-SNOM signals by more than one order of magnitude, while still offering a spatial resolution of about 100 nm at a wavelength of $\lambda$ = 119 μm. On the other hand, exploiting the increase of s-SNOM signals with tip length, we succeeded in sub-15 nm (<$\lambda$/8000) resolved THz imaging employing a tungsten tip with 6 nm apex radius. We explain our findings and provide novel insights into s-SNOM via rigorous numerical modeling of the near-field scattering process. Our findings will be of critical importance for pushing THz nanoscopy to its ultimate limits regarding sensitivity and spatial resolution.**

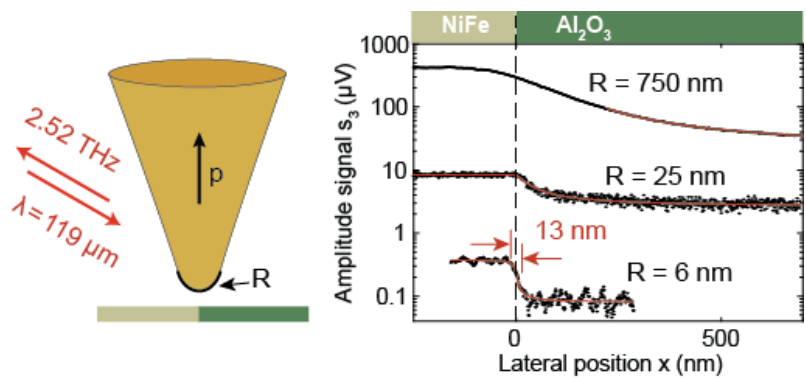

**TOC graphics**

## Introduction

Scattering-type scanning near-field optical microscopy (s-SNOM)[1] is an emerging scanning probe technique, which extends the power of optical techniques deep into the subwavelength regime for nanoscale imaging and spectroscopy from visible to terahertz frequencies. Applications include nanoscale chemical materials identification via molecular vibrational spectroscopy,[2-5] mobile carrier mapping in semiconductor nanostructures,[6-9] polariton mapping in 2D materials,[10-13] and studies of metal-insulator transitions and strongly correlated quantum materials.[14-17]

In s-SNOM, a metallic, cantilevered atomic force microscopy (AFM) tip is illuminated with focused laser radiation and the backscattering from the tip is detected by a far-field detector. The tip acts as an antenna and concentrates the incident radiation to a nanoscale near-field spot (nanofocus) at the very tip apex [18]. When the tip is brought in close proximity to a sample surface, the near-field interaction between tip and sample modifies the tip-scattered field, depending on the local dielectric sample properties. By collecting the tip-scattered field with a distant (far-field) detector, information about the local sample properties close to the tip apex are obtained. Recording the tip-scattered field as a function of tip position, while scanning the sample, yields nanoscale-resolved images of the dielectric sample properties. The spatial resolution is determined essentially by the tip radius, independent of the wavelength. Unavoidable background scattering can be suppressed by vertical tip oscillation and demodulation of the detector signal at a higher harmonics (multiples) of the tip´s oscillation frequency. Improved background suppression is obtained in combination with interferometric detection, which additionally yields amplitude and phase of the scattered field that are related to the reflection and absorption properties of the sample, respectively[19-21].

s-SNOM at THz frequencies[22,23,6,24-33] a tool of rapidly growing interest, as it allows, for example, for THz nanoimaging of complex electronic phases in 2D materials and exotic conductors [34.] However, THz s-SNOM is still in its infancy, mainly due to two reasons. First, the power of most THz sources is rather weak compared to that of infrared and visible sources, and detectors are often bulky, expensive and require cryogenic operation. Second, standard AFM tips have a typical length of 10 to 20 µm, which is subwavelength-scale at THz frequencies. The antenna performance of these tips (i.e. the near-field to far-field scattering efficiency) at THz frequencies is thus much worse than at infrared or visible frequencies. On the other hand, the success of s-SNOM at infrared and visible frequencies can be attributed to a large degree to its readily available and relatively easy-to-use AFM technology, particularly to the possibility of employing cantilevered AFM tips. Thus, design and optimization of cantilevered AFM tips regarding their performance as THz near-field probes will be of crucial importance for the further development of THz s-SNOM.

Key parameters for the performance of AFM tips as infrared and THz near-field probes are tip length L and tip apex radius R, which are expected to determine the near-field scattering efficiency and spatial resolution, respectively. Systematic studies and clear design rules, however, have not been reported so far. Only recently, we demonstrated via photocurrent measurements that the near-field intensity at the tip apex can be enhanced by nearly one order of magnitude by increasing the tip length to match geometric antenna resonances[37]. Yet, it still has to be verified how much the increased near-field enhancement improves the s-SNOM signals, i.e. the near-field scattering efficiency. Furthermore, while it is widely accepted that sharp tips yield better spatial resolution, the relationship between the tip radius and s-SNOM signal spark some controversy. Recent studies with ultra-sharp tips have shown that IR s-SNOM signals nearly vanish for small tip radii[38] despite the larger field enhancement at their tip apex [18] that is considered to be a key factor for enhancing the s-SNOM signal, at least at visible frequencies [39].

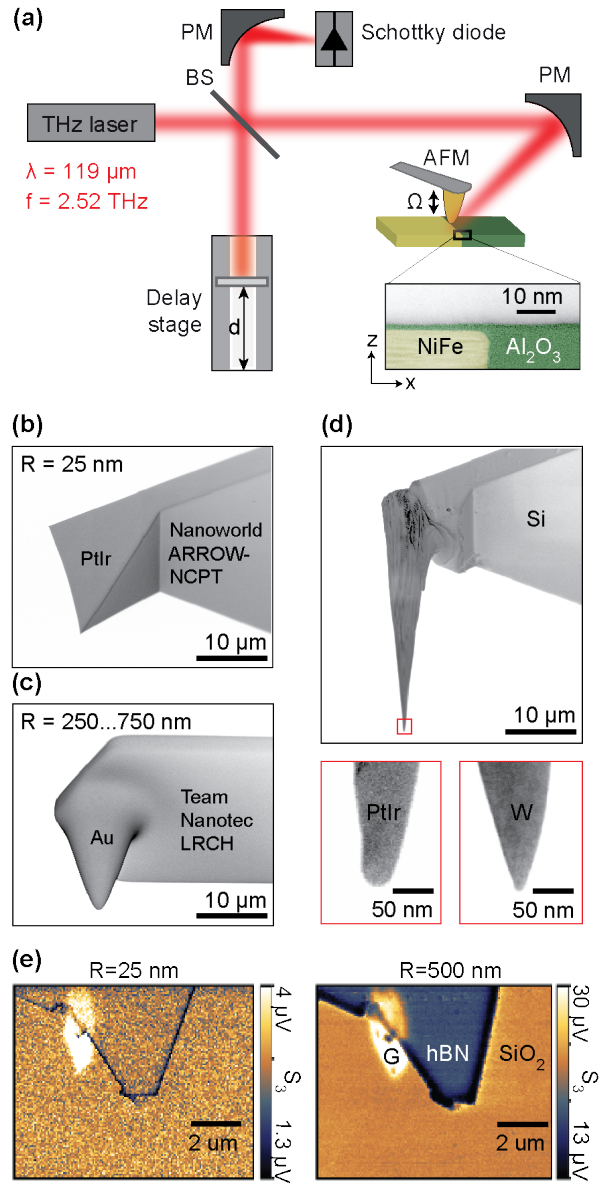

**Figure 1: Setup, tips and imaging examples.** (a) Schematic of the THz s-SNOM setup with Schottky diode as a detector. PM and BS stand for parabolic mirror and beam splitter (double-sided polished silicon wafer), respectively. Inset shows a false-colored Scanning Transmission Electron Microscopy (STEM) image of the read/write head of a HDD serving as a topography-free test sample for evaluating tips performance. (b, c) SEM images of commercial tips used for the radius variation study. (d) SEM of homemade long tips with R = 25 nm and R = 6 nm apex radius made from Platinum-Iridium and Tungsten wire respectively. (e) THz s-SNOM images $s_3$ of a 50 nm thick h-BN flake next to a small graphene flake on top of a SiO$_2$ substrate obtained with a $R = 25\ nm$ (left) and a $R = 500\ nm$ (right) tip. Tapping amplitude was 45 nm.

Here, we systematically study the influence of the tip apex radius on the s-SNOM signal amplitude using commercially available AFM tips with the potential of wide spread use. We find that the signal increases by orders of magnitude with increasing apex radius, while the spatial resolution degrades only moderately. In order to explain these rather unexpected observations, we perform quantitative numerical modelling considering realistic parameters for the tip geometry. Using full wave simulations, we calculate the tip-scattered far-field radiation, which is the quantity that is actually detected in the experiment. Taking into account tip oscillation and signal demodulation in s-SNOM, we obtain unprecedented understanding of the near-field scattering mechanism, background suppression by higher harmonics demodulation and measured near-field signals. Further, we study experimentally the influence of the tip length on the s-SNOM signal. Finally, using a custom made 40 μm long cantilevered AFM tip with an ultrasharp apex, we are able to push the resolution in THz-s-SNOM below 15 nm (<λ/8000) at an imaging wavelength of λ = 119 μm.

**Description of setup, sample and tips**

In order to compare the performance of various tips, we used a THz s-SNOM as sketched in Figure 1a. The setup is based on a commercial AFM-based s-SNOM (Neaspec GmbH, Germany), where the AFM tip is illuminated by a tunable THz gas laser (SIFIR-50, Coherent Inc., USA) operating at 2.52 THz (λ = 119 μm). The light scattered from the tip is recorded interferometrically with a GaAs-based Schottky diode (WR-0.4ZBD, Virgina Diodes Inc. USA). Due to THz absorption by water molecules in the air, as well as due to reflection and diffraction losses at optical elements (we use rather small optics compared to wavelength and beam diameter), the total power reaching the diode is about 1 mW, although the emitted laser power was in the range of 10 mW. Operating in tapping mode (where the tip is oscillating vertically at the cantilever´s mechanical oscillation frequency Ω) leads to a signal modulation at Ω and higher harmonics at nΩ (n>1), since the near-field signal depends nonlinearly on the tip-sample distance. By demodulation of the detector signal at higher harmonics with n>2 we can suppress additive background scattering contributions. Moreover, the interferometric detection scheme allows for amplitude- and phase-resolved near-field imaging, as well as for suppression of multiplicative background scattering contribution [20]. Since our sample does not show any significant phase response, we fix the reference mirror (mounted on a delay stage) at the position of maximal constructive interference between the reference field and tip-scattered field. In this case, we directly detect the signal amplitude $s_n$ (where the index n indicates the demodulation order).

Upon tip replacement the new tip position might differ slightly from the old one, resulting in suboptimal focusing and detection of the THz beam. Therefore, after each tip replacement, the parabolic mirror, the reference arm mirror and the detector position were realigned to maximize the signal amplitude, while the laser emission power was kept constant. The employed tapping amplitude was about A = 50 nm and demodulation order n = 3 unless otherwise noted.

As a test sample we use a commercial hard disk read/write head, more specifically the well-defined sharp metal/dielectric interface therein (the inset of Figure 1a shows a transmission electron micrograph of a cross-section of the sample)[38]. Recording line scans across the sharp material interface allows for simultaneous determination of the absolute amplitude signal $s_n$, the signal contrast between the two materials, and the lateral resolution.

In order to study the influence of the tip radius on the absolute signal amplitude we used commercially available AFM tips. The R = 25 nm tip is a PtIr-coated silicon tip (model ARROW-NCPT, Nanoworld) with a pyramidal shaft. The tips with larger radii of $R$ = 250, 500 and 750 nm are gold coated silicon tips (model LRCH, team nanotec) and have a conical shape. Scanning electron microscopy (SEMs) images of the tips with $R$ = 25 nm and $R$ = 500 nm are shown in Figure 1b. To study the influence of tip length variations, we fabricated tips from Pt80/Ir20 (in the following referred as to Pt/Ir) and W [37] wires by focused ion beam (FIB) machining. Figure 1c shows an SEM image of a whole tip, as well as the apex of a PtIr tip with $R = 25$ nm and of a W tip with $R = 6$ nm. The later was applied to obtain THz images with an unprecedented spatial resolution better than 15 nm.

Before describing our systematic study employing the test sample, we demonstrate in Figure 1e the dramatic influence of the tip radius with an imaging example. We imaged a sample consisting of a 50 nm thick h-BN flake next to a small graphene flake on top of a SiO$_2$ substrate with tips of apex radius $R = 25$ nm (left) and $R = 500$ nm (right). We clearly see the enormously enhanced signal-to-noise-ratio and material contrast for the lager tip radius, while the resolution seems to be diminished comparatively little. The contrast between graphene and the SiO$_2$ substrate stems from the metallic conductivity of graphene [35]. More important, the comparatively weak contrast between h-BN and SiO$_2$ can be recognized reliably only with the R = 500 nm tip. It reveals that in this specific sample the h-BN exhibits a slightly smaller permittivity than SiO$_2$ (note that for dielectrics with small positive permittivity the s-SNOM amplitude signals s$_n$ increase monotonously with increasing permittivity[40,41]). At the h-BN edges we see for both tips a dark rim, which is wider for the larger tip. To avoid this well-known edge-darkening artefact [41] (which obviously increases with tip radius and can lead to an overestimation of resolution) in the evaluation of the performance of the various tips, we have chosen to use the ultra-flat well-defined test sample described above.

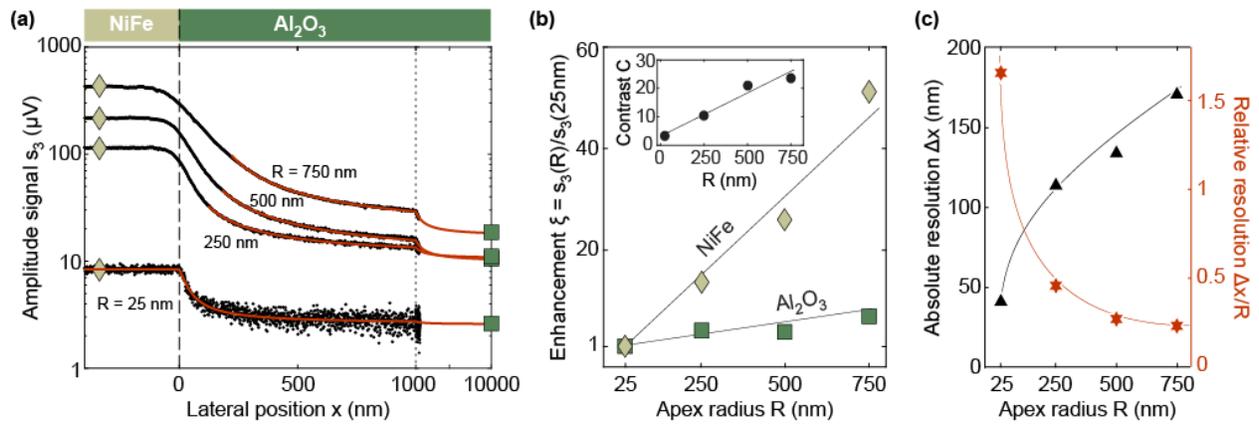

**Figure 2: s-SNOM measurements across the well-defined metal/dielectric interface of the HDD read/write head.** (a) THz near-field scattering amplitude s$_3$ line profiles (dots) obtained with four different commercial tips with apex radii of 25 nm, 250 nm, 500 nm, and 750 nm, respectively. The data are plotted on a logarithmic scale. Solid lines are fits using the integral of an asymmetric Lorentzian as described in the Supporting Information S1. (b) Signal amplitude on metal and dielectric normalized to the signal of the 25 nm tip, respectively, s$_3$(R)/s$_3$(25nm). Inset shows the contrast C = s$_3$(metal)/s$_3$(dielectric). (c) Resolution Δx and relative resolution Δx/R determined by via the FWHM of the derivative of the line profiles in Figure 2(a).

**Tips with large apex yield larger amplitude signals at the cost of an only moderate decrease of resolution**

In Figure 2a we show $s_3$ line profiles across the NiFe/Al$_2$O$_3$ boundary of the test sample recorded with tips of different apex radii. Interestingly, we observe an overall increase of the amplitude signal by more than one order of magnitude when increasing the tip apex radius from 25 nm to 750 nm. This trend is contrary to the general view that the amplitude signal should scale with the field enhancement at the tip apex, which increases with decreasing apex radius. We will discuss this aspect below when discussing the numerical simulations. For a quantitative comparison, we introduce the signal enhancement factor with respect to the standard R = 25 nm tip, $\xi = \frac{s_3(R)}{s_3(25\text{ nm})}$. As follows from Figure 2b, the signal enhancement is not the same for metal and dielectric parts, reaching an impressive factor of 50 on the NiFe (metallic) surface and 7 on Al$_2$O$_3$ for the tip with $R$ = 750 nm. Apparently, the signal enhancement is stronger for materials with large near-field reflection coefficient $\beta = (\varepsilon_s - 1)/(\varepsilon_s + 1)$ [42]. The material-dependent signal enhancement leads in consequence to a material contrast $C = s_{3,Metal}/s_{3,Dielectric}$ that is dependent on the apex radius $R$, as shown in the inset in Figure 2b. This contrast enhancement is not unexpected, as the increase in radius reduces the relative tapping amplitude $A/R$, which is known to enhance the material contrast [19,43]. Nevertheless, this is the first time an impressive 30-fold contrast enhancement is reported.

Another trend observed in the $s_3$ line profiles in Figure 2a is the broadening of the material interface with increasing tip radius, corresponding to an expected degradation of spatial resolution when the tip radius increases. We also note that the line profiles across the metal-dielectric boundary are asymmetric, which can be explained by a material-specific screening of the tip´s near fields being stronger on the metal than on the dielectric.[38] Following the approach developed in Mastel et al[38], we quantify the lateral resolution $\Delta x$ as the full width at half maximum (FWHM) of the derivative of the $s_3$ line profiles. We find that the resolution decreases sub-proportional with apex radius (Figure 2c). When increasing the tip apex radius by a factor of 30, the resolution decreases only by a factor of 4 (from 41 nm to 170 nm). We explain the surprisingly good lateral resolution of tips with large apex radii by the extremely small relative tapping amplitude $A/R \ll 1$, yielding to an extreme virtual tip sharpening effect [43]. With standard tips of $R = 25$ nm such a regime would require a tapping amplitude $A$ of only a few nanometers. The orders-of-magnitude increase in signal provided by large tips, in conjunction with their capability of still providing a spatial resolution on the 100 nm scale, could highly benefit applications that require high sensitivity but not uttermost spatial resolution.

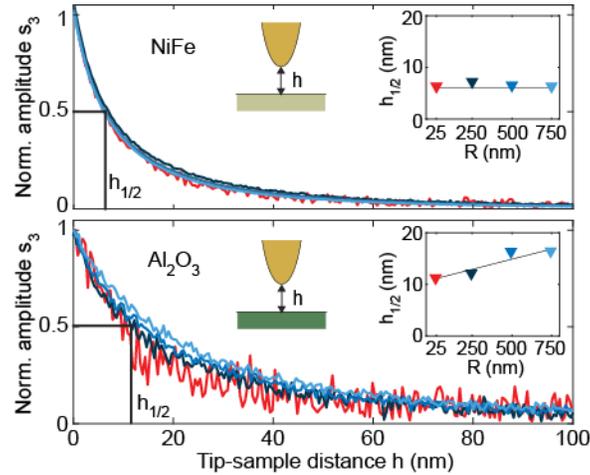

**Figure 3** : Approach curves for tips with different tip radii recorded on the metallic NiFe (top) and dielectric Al₂O₃ (bottom). Insets show the distance $h_{1/2}$ at which the signal drops to half of its maximum.

**Approach curves confirm background-free near-field signals**

In order to verify that we measure exclusively near-field signals (i.e. that we can exclude background contributions), we recorded approach curves for each tip and on each sample material, that is, we measured the amplitude signal $s_3$ as a function of the tip-sample distance *h* on NiFe and Al₂O₃ (Figure 3). We observe that $s_3$ rapidly decreases with increasing distance *h* and asymptotically approaches zero, thus confirming background-free near-field signals. Most surprisingly, we find that the decay distance, $h_{1/2}$, at which the amplitude signal has decayed by a factor of two, is nearly independent of the apex radius *R*. On NiFe, we measure $h_{1/2} = 6$ nm for all tips. On Alumina, $h_{1/2}$ increases only by a factor of 1.5 (from 11 to 16 nm) when the tip radius increases by a factor of 30 (from R = 25 nm to 750 nm). These dramatic sub-radius decay distances are in stark contrast to the widespread assumptions[6,19,27,39,43] that (i) the near-field signal´s decay distance is on the scale of the tip radius, and (ii) the lateral resolution is closely related to $h_{1/2}$. Comparison of Figure 3 and Figure 2c reveals that $h_{1/2}$ can be one order of magnitude smaller than the lateral resolution Δ*x* (e.g. $h_{1/2}$ = 16 nm << Δ*x* = 170 nm for the tip with R = 750 nm). For that reason, generally, the decay distance measured from approach curves can neither be used as an estimate for the lateral resolution nor as measure of the tip radius. We explain the rapid amplitude signal decay similarly to the surprisingly good lateral resolution of tips with large apex radii: The relative tapping amplitude $A/R \ll 1$ is extremely small, yielding to an extreme virtual tip sharpening effect [43] that would occur upon operating a standard tip with $R = 25$ nm with tapping amplitudes of only few nanometers or below.

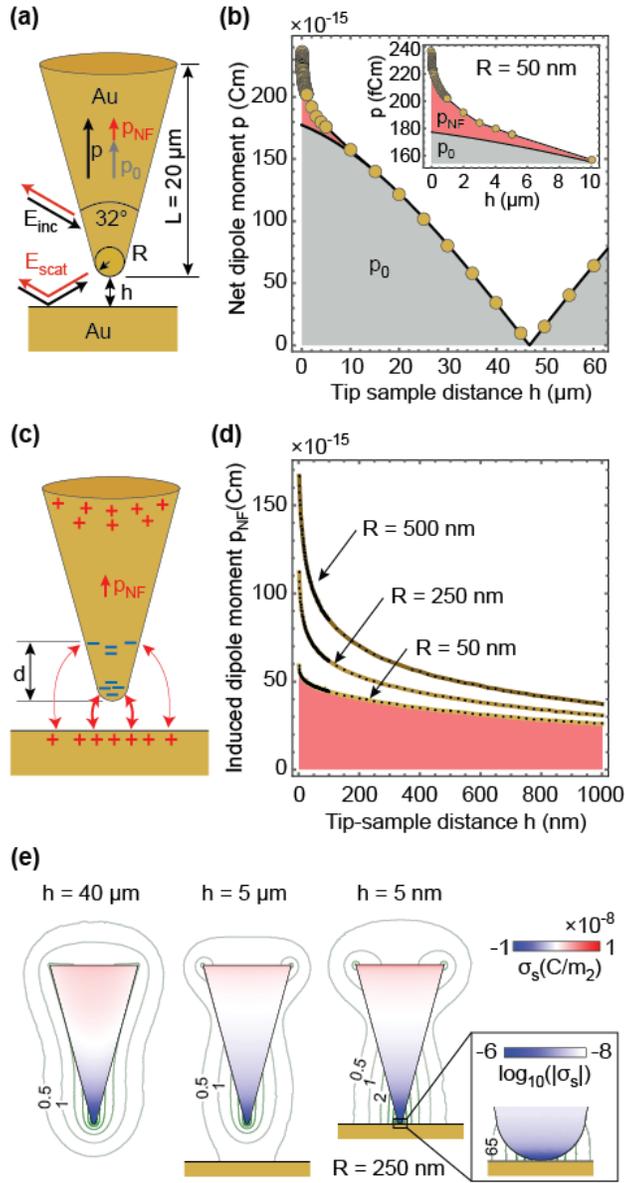

**Figure 4: Numerical modeling of near-field induced scattering from the tip.** (a) Sketch of the numerical model. (b) Calculated total dipole moment p (dots), illumination-induced dipole moment $p_0$ (black line and grey area) and near-field interaction induced dipole moment $p_{NF}$ (red area). Inset shows a zoom-in to small tip-sample distances. (c) Schematics describing the physical interaction process. (d) Calculated dipole moment $p_{NF}$ for three different apex radii R. (e) Simulated surface charge distribution $\sigma_s$ (color map) and electric near-field distribution $f_{NF}$ (contour plot, logarithmic scale) for the tip above a gold surface (normalized to $p_o$). The number on the outermost contour indicates the value of $|f_{NF}|$. Right panel shows a zoom of the apex region (logarithmic color map).

**Numerical full-wave simulations confirm the observed experimental trends**

In order to understand the increasing amplitude signal $s_3$ with increasing tip apex radius $R$ and why the vertical signal decay length $h_{1/2}$ is independent of $R$, we performed full wave numerical simulations (COMSOL) of the scattering by a tip with realistic dimensions. Geometric considerations are essential for a proper description of s-SNOM signals[44] since the established analytical models [1, 45,46] can provide only image contrasts rather than signal strengths.

The tip geometry used for our simulations is illustrated in Figure 4a and closely matches that of the commercial tips shown in Figure 1c. The tip-sample system is illuminated by an external illuminating field $E_{inc}$, which is modeled as a plane wave incident at 30° to the sample surface and its partial reflection at the sample surface. The illuminated tip interacts with the sample via the near fields around the tip apex, resulting in a net vertical dipole $p$ (the horizontal dipole is negligible, see Methods). The tip-scattered electric field amplitude, $E_{scat}$, can be considered as the radiation of this net dipole (since the tips of about 15 µm length are much shorter than the THz wavelength of 119 µm), and is thus proportional to $p$. The latter can be evaluated as the first moment of the surface charge density, $\sigma_S$, induced in the tip by both the illumination and near-field interaction between tip and sample. Subsequently we obtain the amplitude of the tip-scattered field by

$$E_{scat} \propto |p| = \left| \iint_S \sigma_S z \, dS \right| \qquad (1)$$

where d$S$ is the surface element and $z$ its vertical position (for details see Methods Section).

First, we analyze the tip-sample interaction for a tip that is statically placed at a distance $h$ above the sample. Figure 4b shows the calculated net dipole moment $p$ for a tip with apex radius $R = 50$ nm as function of distance $h$ between tip apex and a semi-infinite gold sample. $p$ is dominated by a standing wave pattern with a periodicity of $\sim\lambda/2 = 59$ µm (gray area in Figure 4b). We identify a standing wave pattern caused by the illumination-induced dipole moment $p_0$, which is generated purely in response to the illuminating field $E_{inc}$ (i.e. in absence of tip-sample near-field interaction). Its spatial variation results from the interference of the direct incident wave and the wave reflected from the sample surface (see Supporting Information S2). For small tip-sample distances $h < \lambda/10$ we observe a strong increase in $p$ (red shaded area) on top of the standing wave pattern. The corresponding additional dipole moment $p_{NF} = p - p_0$ (red shaded area) results from the near-field interaction between the tip and the sample and contains information on the local optical sample properties. A closer look at $p_{NF}$ in the range $h < 10$ µm ($\sim\lambda/10$) (see inset of Figure 4b) reveals two distinct near-field interaction regimes: (i) a slowly increasing near-field interaction for $h \gg R$ (long-range regime) that has been barley observed and described yet, and (ii) a strong and rapid increase of the near-field interaction for $h \rightarrow 0$ (short-range regime) that is well known from s-SNOM experiments and modelling.

In order to understand the long-range near-field interaction, we resort to the model depicted in Figure 4c. The charges arising from the illumination-induced dipole $p_0$ interact with the sample [46] and generate the dipole moment $p_{NF}$ via near-field interaction. The interaction strength (depicted by red arrows in Figure 4c) of a charge located at a distance $d$ from the apex, and thus the amount of near-field-induced charge $\sigma_{S,NF}$, decreases quickly as function of its distance from the sample surface, $d + h$. For $d \gg R$, the interaction strength varies slowly as function of tip-sample distance $h$ (since $d + h \approx d$), leading to a slow long-range increase of $p_{NF}$ when $h$ decreases. This interpretation is supported by the full-wave simulations shown in Figure 4e, where we plot the electric near-field distribution, $E_{NF} = a * (E - E_{inc})$ with $a = p_0(h = 40\mu m)/p_0(h)$ being a normalization factor, and, $\sigma_S$ being the induced surface charge density. We clearly see that at a tip-sample distance of $h = 40\ \mu m$ the tip's near-field is not yet disturbed by the presence of the sample. At $h = 5\ \mu m$, however, the near field already becomes distorted due the sample, which confirms the presence of the long-range near-field interaction regime described above. The strength and onset are essentially determined by the overall tip geometry (i.e. tip opening angle and length) but not by the apex radius, thus leading at $h \gg R$ to a nearly radius-independent dipole moment $p_{NF}$ that slowly decreases with increasing $h$ (compare approach curves for different apex radii in Figure 4d. Interestingly, we do not observe long-range near-field signals in the experimental approach curves shown in Figure 3, which will be discussed and resolved below once we consider the tip-sample distance modulation and signal demodulation.

More important, in the short-range near-field interaction regime, *i.e.* for $h < R$, we observe a much stronger and more rapid increase of the near-field-induced dipole moment $p_{NF}$ (Figure 4d), which arises from the interaction of charges that are located at the very tip apex. The interaction of these charges increases rapidly with decreasing tip-sample distance $h$ (since $d < R$ and thus $d + h \approx h$), and so does the resulting dipole moment $p_{NF}$.

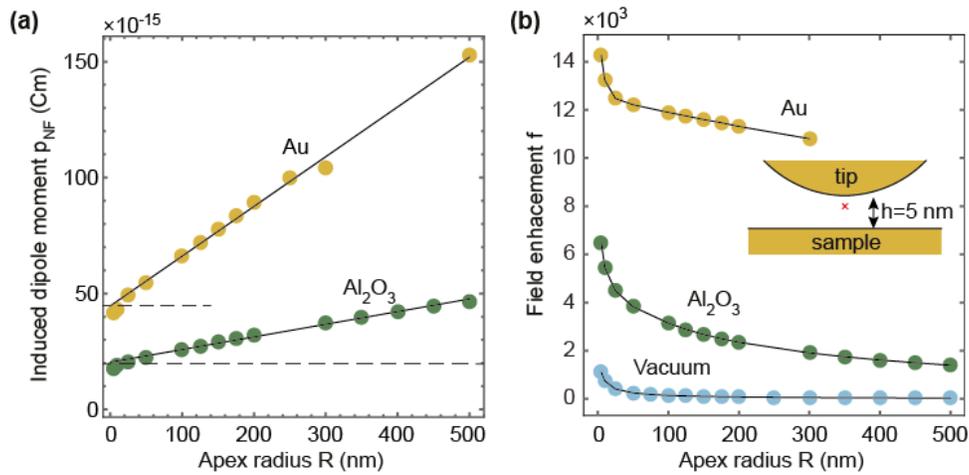

**Figure 5: Near-field scattering versus near-field enhancement.** (a) Simulated near-field-induced dipole moment $p_{NF}$ evaluated for a tip-sample distance of $h = 5\ nm$ for NiFe and Al$_2$O$_3$ samples. The dashed lines indicate the dipole moment induced by the long-range near-field interaction, i.e. the contribution of the tip shaft to the tip-sample near-field interaction. Tip geometry and size are the same as in Figure 4a. (b) Field enhancement (normalized for the tip volume) evaluated $1\ nm$ below the tip apex (position marked by red cross in inset), under the same conditions as in Figure 5a. Additionally, the field enhancement for an isolated tip is shown (vacuum).

For quantifying the influence of the apex radius on the near-field-induced dipole moment $p_{NF}$, we performed calculations for various apex radii. In Figure 4d we observe that $p_{NF}$ strongly increases with increasing apex radius $R$ in the limit $h \to 0$, reproducing the same trend as observed experimentally for the demodulated amplitude signal $s_3$. Systematic simulations for a wide range of apex radii on gold and Al$_2$O$_3$ samples reveal a linear increase of $p_{NF}$ with increasing tip radius (Figure 5a). The slope depends on the sample material, which further confirms the experimentally observed trends. For comparison, we show in Figure 5b the near-field enhancement $f$ between tip apex and sample surface. We observe the contrary behavior, that is, the well-known increase of the field enhancement with decreasing tip radius. Our simulations thus clearly confirm that the increasing field enhancement at a smaller tip apex does not lead to an increased near-field scattering. Apparently, the decreasing field enhancement for larger apex radii is more than compensated by the larger near-field interaction area that leads to an increased capacitive coupling when $R$ increases, thus to an increase of $p_{NF}$.

We note that our findings show a clear similarity to findings made in scanning electrostatic force microscopy (EFM), where an increase of the tip radius leads to stronger EFM signals owing to an increased capacitive coupling between tip and sample.[47-50] In the future it might be interesting to elaborate whether analytical EFM models [48,49,51,52] could be adapted for quantitative modeling of THz s-SNOM utilizing tips of sub-wavelength scale length (i.e. operating in the non-retarded quasi-electrostatic limit).

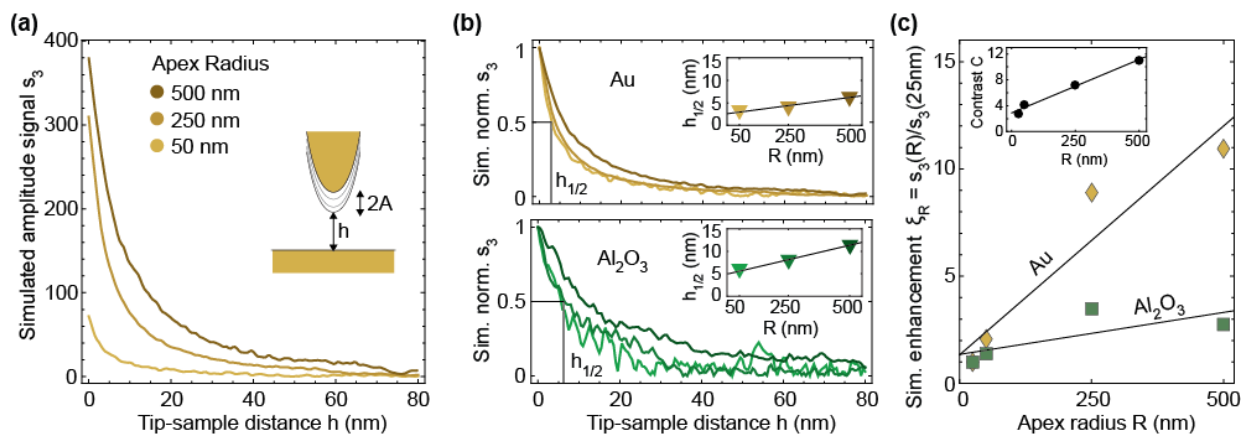

**Figure 6: Numerical simulations considering tip modulation and signal demodulation.** (a) Simulated $s_3$ approach curve for different tip apex radii. Sample is gold. (b) Simulated normalized approach curves for Gold (top) and Alumina (bottom) samples. Insets show the distance at which the signal decreases to half maximum (solid lines are linear fits). (c) Simulated signal enhancement for Gold and Alumina samples. Inset: contrast calculated from simulation (solid lines are linear fits).

**Demodulation of scattered field leads to enhanced near-field sensitivity**

In the previous section we found that the simulation reproduces the experimentally observed trends, essentially that the scattered field (proportional to $p_{NF}$) induced by the near-field interaction between tip and sample increases with increasing tip radius. However, the near-field-induced scattered field does not decay to zero on a sub-100 nm scale, which is observed in the s-SNOM experiments (compare Figure 3 and 4d). The long-range signal decay in the simulations results from the discussed shaft-sample near-field interaction and is nearly insensitive to small variations of the tip height (see Figure. 4e). Thus, we expect it to be suppressed by the higher-harmonics demodulation procedure utilized in the experiment. In order to recover the experimental results, we thus need to consider tip modulation and signal demodulation in our simulations. Mathematically, the demodulated signal can be described as the n-th Fourier coefficient of the height-dependent scattered field $E_{scat}[h(t)]$ with $h(t) = A(1 + \cos(\Omega t))$:

$$s_n = \int_{-\infty}^{\infty} E_{\text{scat}}(h(t)) e^{in\Omega t} dt \qquad (2)$$

Since $E_{scat}$ is proportional to $p$ (eq. (1)), the demodulated signal is proportional to the Fourier coefficient of $p(h(t))$, which we obtain by interpolating the simulated net dipole. Indeed, after demodulation at the third harmonic, i.e. calculating $s_3$, we recover a strong signal drop on the 10 nm scale when the tip-sample distance is increased (see Figure 6a). Moreover, the signal drops asymptotically to zero, as observed in the experiment. For a direct comparison with the experimental approach curves shown in Figure 3, we plot in Figure 6b the normalized approach curves on gold and $Al_2O_3$. We find a good quantitative agreement: (i) the approach curves decay on both materials and for all tips radius within less than 20 nm and (ii) the decay length $h_{1/2}$ depends more strongly on the apex radius when the tip is placed above the $Al_2O_3$ sample. Further, in Figure 6c we find good agreement between the experimental and simulated signal enhancement $\xi$ and material contrast C. The quantitative discrepancy between experiment and simulation might arise from multiple factors related to the experiment. In the experiment, the smallest tip with the smallest radius (to which the measurements are normalized) has a different geometry and metal coating compared to the tips with larger apex radius (pyramide with PtIr coating vs. cones with Au coating, see Figure 1). Further, in the simulation we do not consider the cantilever, variations in tip volume, as well as the thin (1 nm) $Al_2O_3$ layer covering NiFe in the real sample. Altogether, our numerical simulations including tip modulation and signal demodulation clearly verify the increasing s-SNOM signals for tips with larger apex radius, and that the signal decay length $h_{1/2}$ of approach curves is nearly independent of the tip radius.

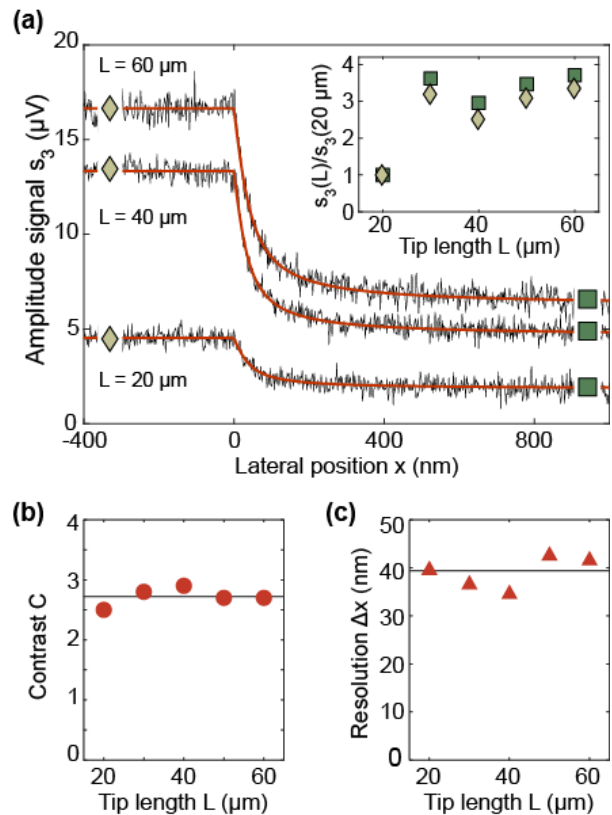

**Figure 7: s-SNOM measurements across the well-defined metal/dielectric interface of the HDD read/write head.** (a) Line profiles recorded with custom-made PtIr-tips of different length L, all with apex radius R=25 nm. Inset shows the amplitude signal $s_3$ on the metal and dielectric areas, respectively, normalized to the 20 µm long tip. (b) Material contrast C as function of tip length L. (c) Lateral resolution $\Delta x$ as function of tip length L. The horizonal lines in (b) and (c) indicate the average value.

**Long tips enhance THz s-SNOM signals**

Increasing the tip radius is not suitable in cases where both high sensitivity and highest spatial resolution are required. In a recent work [37], we showed that the field enhancement at the tip apex can be increased by increasing the tip length to several tens of micrometers, while keeping the tip apex radius constant. In these experiments, however, we measured the photocurrent produced by the tip´s near field with a graphene-based photodetector, rather than standard s-SNOM signals. For that reason, we study in the following the influence of the tip length $L$ on the amplitude signal $s_3$, material contrast $C$, and lateral resolution $\Delta x$. We fabricated six PtIr tips with an apex radius $R = 25$ nm and lengths $L$ spanning from 20 µm to 60 µm. Three line profiles are shown in Figure 7a. We observe a trend to larger amplitude signals $s_3$ for tips of increasing length $L$. The amplitude signal enhancement $\xi_L = \frac{s_3(L)}{s_3(20\ \mu m)}$ (inset to Figure 7a) reaches almost a factor 4. This value is in good agreement with our previous photocurrent measurements, which observed an *intensity* enhancement by factor 9. We attribute the absence of a clear antenna resonance of the tip for $L\sim\lambda/2$ to experimental uncertainty due to alignment of the interferometer. Material contrast $C$ (Figure 7b) and lateral resolution $\Delta x$ (Figure 7c) are both independent of the tip length within experimental accuracy. This finding indicates that the tip length does mainly influence the efficiency of coupling near-field interaction and far-field radiation, but has no significant influence on the tip-sample near-field interaction itself. This situation might change if the material exhibits a strong resonance at the illumination wavelength (*i.e.* phonon or plasmon resonances). In that case, the coupling between tip resonances and material resonances might lead to intricate spectral modifications of the s-SNOM amplitude signal.

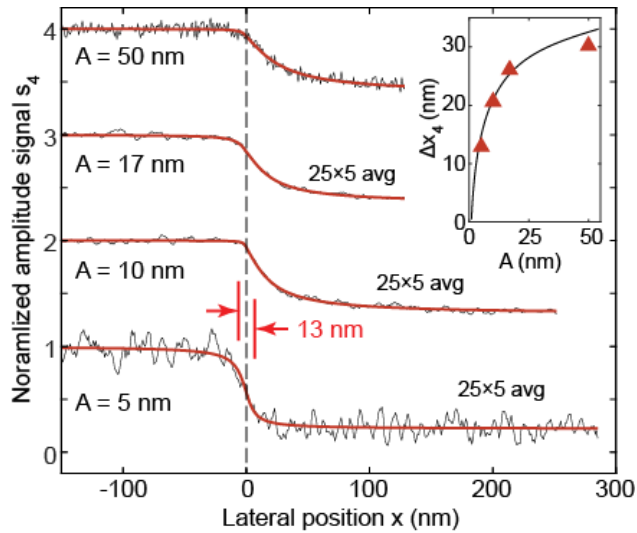

**Figure 8: Pushing the resolution with sharp and long tips.** Line profiles across the metal/dielectric boundary of the test sample recorded with a W-tip with R = 6 nm at varying tapping amplitude *A* (avg: number of averaged scans of which each one was smoothened with an average filter over 5 adjacent data points). Inset: Resolution Δx as function of tapping amplitude.

**Sub-15 nm resolution achieved with ultrasharp tips**

Finally, we aim to push the lateral resolution $\Delta x$ in THz s-SNOM by fabricating long and ultrasharp tungsten tips, combining the prerequisites for efficient conversion of near-field interaction into far-field scattering and for high resolution imaging, respectively. Figure 1d shows a tip with $R = 6$ nm and $L = 40$ µm, which was used to record the line profiles across the material boundary of our test sample. We applied signal demodulation at the 4$^{th}$ harmonic, yielding $s_4$, in order to enhance the virtual tip sharpening effect. The $s_4$ line profiles are shown in Figure 8 for a series of tapping amplitudes decreasing from $A = 50$ nm to 5 nm. In order to achieve a reasonable signal-to-noise-ratio (SNR > 1) for the smallest tapping amplitude, we averaged 25 line scans (See Supporting Information S3) and performed a moving average filter over 5 points (operating on the complex-valued data), leading to an effective integration time of 3.25 s per point and a slight degradation of the lateral resolution. We observe a clear improvement of the resolution for very small tapping amplitudes $A \lesssim R$ (inset of Figure 8). A lateral resolution of $\Delta x = 13$ nm could be achieved for the smallest tapping amplitude of $A = 5$ nm. An even better spatial resolution could in principle be achieved by using smaller tapping and/or higher demodulation orders, being limited, however, by a further decreasing SNR.

**Remarks**

Throughout this study we worked in a regime of extremely subwavelength radii $R < \lambda/100$. For larger tip radii we expect a cut-off for the signal enhancement $\xi_R$ due to retardation effects. Indeed, earlier experiments with passive s-SNOM operating at $\lambda = 15$ µm have shown a negligible effect of the apex size on the near-field signal [53], and our preliminary observations with s-SNOM in the mid-infrared frequency range ($\lambda \sim 10$ µm) indicate a saturation of the signal enhancement for apex radii larger than 100 nm. Indeed, an apex radius of 750 nm at 2.52 THz corresponds to a radius of 50 nm or 2.5 nm when scaling the experiment to mid-infrared or visible frequencies, respectively. Our insights will therefore be relevant to estimate the effect of extremely sharp tips at visible and mid-infrared frequencies when aiming at sub-nm or atomic scale lateral resolution.

We also note that in this work we studied the origin of the s-SNOM signals only for extended material domains with dimensions much larger than $R$. Future studies are required to see how the s-SNOM signal depends on the apex radius when isolated particles with dimensions smaller than $R$ are imaged.

*Conclusions*

We performed systematic experiments and numerical simulations to study the near-field signals in THz s-SNOM as a function of the tip apex radius when standard AFM tips are employed. The simulations considered a realistic tip geometry, as well as the signal processing procedure, i.e. tip-sample distance modulation and signal demodulation at higher harmonic of the modulation frequency. In good agreement between experiments and theory, we found that the THz s-SNOM signal in general does not scale exclusively with the electric field enhancement at the tip apex. s-SNOM signals and field enhancement even show opposite trends as function of the apex radius. Particularly, by increasing the radius of an AFM tip (standard length of 20 µm) from 25 nm to 750 nm, we could increase the THz near-field signal by more than one order of magnitude, while the resolution degrades by only less than a factor of 4 (still being in the range of 100 nm). In cases where high sensitivity but not a maximum resolution is required, standard AFM tips with large apex radius thus provide a readily available solution to boost the near-field signals in THz s-SNOM applications.

Our numerical study clearly shows that the field enhancement at the tip apex is not sufficient to predict the strength of s-SNOM signals. Instead, the conversion of the near-field interaction between tip and sample into far-field scattering (the quantity actually measured in s-SNOM and described in this work by the net dipole moment of a tip with realistic dimensions) has to be considered to correctly predict the strength of s-SNOM signals.

In a second series of experiments we performed THz s-SNOM with custom-made AFM tips of increased tip length but constant tip apex radius of about 25 nm. The tips of several micrometer length increase the s-SNOM amplitude signal by a factor 4, while nm-scale spatial resolution is maintained. The increase in tip length leads to increased far-field coupling efficiency. At the same time, by keeping the tip shaft and apex constant, we showed that the near-field tip-sample interaction is not influenced by the tip length. Finally, using a custom-made tip with $R = 6$ nm and an extremely small taping amplitude of $A = 5$ nm, we achieved a resolution in the sub-15 nm range, although for practical applications the SNR has to be further improved by more sophisticated tip engineering in the future.

*Acknowledgments*

The authors acknowledge support from the Spanish Ministry of Economy, Industry, and Competitiveness (national project MAT2015-65525 and the project MDM-2016-0618 of the Marie de Maeztu Units of Excellence Program), the European Commission under the Graphene Flagship (GrapheneCore2), the H2020 FET OPEN project PETER (GA#767227), and the Swiss National Science Foundation (Grant No. 172218).

*Competing Interests*

R.H. is a co-founder of Neaspec GmbH, a company producing scattering-type scanning near-field optical microscope systems such as the one used in this study. The other authors declare no competing interests.

## Methods

### Full wave numerical simulations

Our numerical simulations were performed using commercial finite element method (COMSOL Multiphysics) based on solving Maxwell's equations in the frequency domain. The tip was modeled as a conical frustum and a semispherical apex with geometric parameters given in Figure 4a. The influence of the cantilever was neglected. The first 500 nm of tip starting from the apex bottom were modelled as solid Au. The rest of the tip was replaced with impedance boundary conditions in order to save computer memory. The dielectric constants of Au and Alumina were taken as $-710000 + 248000i$ and 9.61, respectively [54] In order to properly describe the tip-sample illumination in relatively small, comparable to wavelength simulation domain (200 μm by 100 μm by 100 μm+ tip-sample distance), the scattering problem approach was chosen with the background electric field $\mathrm{E}_{inc}$ (i.e. in the absence of the tip) defined as a p-polarized direct plane wave incident at 30° relative to the surface and the corresponding reflected wave due to Fresnel reflection from the sample surface.. The surface charge density was calculated as:

$$\sigma_S \sim \iint_S \mathbf{E} \cdot \mathbf{n} dS \qquad (3)$$

where **E** is the electric field vector and **n** is the outward normal of the surface element d*S*. The surface integral in Eq. (1) was carried out over the full tip. Here we only considered vertical (z) component of the dipole moment, as normally done in theoretical description of s-SNOM scattering[46]. This is because the in-plane (x-y) dipole is typically small due to tip being much more elongated vertically than longitudinally, resulting is much larger vertical polarization. In addition, the radiation of in-plane dipole is suppressed (nearly completely for metals) by its image dipole.